\documentclass[dvips]{acta}
\usepackage{supertabular,lscape,epsfig}
\usepackage{amssymb}
\usepackage{amsmath}

\newcommand{\cu}{(241944) 2002 CU$_{147}$}
\newcommand{\cus}{2002 CU$_{147}$}
\newcommand{\gh}{2007 GH$_{6}$}
\newcommand{\sa}{2006 SA$_{387}$}
\newcommand{\ql}{2006 QL$_{39}$}
\newcommand{\Larsen}{200P/Larsen}
\newcommand{\qq}{2001 QQ$_{199}$}
\newcommand{\aee}{2004 AE$_{9}$}
\newcommand{\ar}{P/2002 AR$_2$ LINEAR}
\newcommand{\wc}{P/2003 WC$_7$ LINEAR-CATALINA}
\newcommand{\ars}{P/2002 AR$_2$}
\newcommand{\wcs}{P/2003 WC$_7$}

\newcommand{\hr}{(118624) 2000 HR$_{24}$}
\newcommand{\ug}{2006 UG$_{185}$}

\def\farcs{\hbox{$.\!\!^{\prime\prime}$}}

\SetPages{113}{131}

\SetVol{62}{2012}

\begin{document}

\begin{Titlepage}
\Title{Behavior of Jupiter Non-Trojan Co-Orbitals}



\Author{Pawe\l\  W~a~j~e~r and Ma\l gorzata K~r~\'o~l~i~k~o~w~s~k~a}{Space Research Centre of Polish Academy of Sciences,
Bartycka 18A, 00-716 Warsaw, Poland\\
e-mail: wajer@cbk.waw.pl}



\Received{November 18, 2011}
\end{Titlepage}

\Abstract{Searching for the non-Trojan Jupiter co-orbitals we have numerically integrated orbits of 3\,160 asteroids and 24 comets discovered by October 2010 and situated within and close to the planet co-orbital region. Using this sample we have been able to select eight asteroids and three comets and have analyzed their orbital behavior in a great detail. Among them we have identified five new Jupiter co-orbitals: \cu, \sa, \ql, \gh, and \Larsen, as well as we have analyzed six previously identified co-orbitals: \hr, \ug, \qq, \aee, \wc\ and \ar.

\cu\ is currently on a quasi-satellite orbit with repeatable transitions into the tadpole state. Similar behavior shows \gh\ which additionally librates in a compound tadpole-quasi-satellite orbit. \ql\ and \Larsen\ are the co-orbitals of Jupiter which are temporarily moving in a horseshoe orbit occasionally interrupted by a quasi-satellite behavior. \sa\ is moving in a pure horseshoe orbit. Orbits of the latter three objects are unstable and according to our calculations, these objects will leave the horseshoe state in a few hundred years. Two asteroids, \qq\ and \aee, are long-lived quasi-satellites of Jupiter. They will remain in this state for a few thousand years at least. The comets \ar\ and \wc\ are also quasi-satellites of Jupiter. However, the non-gravitational effects may be significant in the motion of these comets. We have shown that \wcs\ is moving in a quasi-satellite orbit and will stay in this regime to  at least 2500 year. Asteroid \hr\ will be temporarily captured in a quasi-satellite orbit near 2050 and we have identified another one object which shows similar behavior - the asteroid \ug, although, its guiding center encloses the origin, it is not a quasi-satellite. The orbits of these two objects can be accurately calculated for a few hundred years forward and backward.}{Minor planets, asteroids: general - Comets: general - Celestial mechanics - Methods: numerical}

\section{Introduction}
Small bodies locked in a co-orbital region are usually associated with the Lagrangian equilibrium points. For more than one century the Lagrange points were only a subject of theoretical considerations. The first object, Trojan asteroid (588) Achilles, that traces tadpole-shaped trajectory (TP) around Jupiter L$_4$ lagrangian point was discovered by Max Wolf in 1906. Trojan asteroids have also been discovered for Mars (Connors et al. 2005) and Neptune (Sheppard and Trujillo 2006).

Shortly after the discovery of the first Trojan, Brown (1911) indicated, that stable horseshoe (HS) orbits, that surround L$_4$, L$_5$ and L$_3$ Lagrangian points can exist. Another kind of 1:1 resonant trajectories, not associated with the Lagrangian points, recently known as quasi-satellite (QS) orbits (Lidov and Vashkov'yak 1994, Mikkola and Innanen 1997), were predicted by Jackson (1913).

All these families of orbits can be simply classified through the librational properties of the principal resonant angle, $\sigma$, which is defined by $\sigma=\lambda-\lambda_p$, where $\lambda$ and $\lambda_p$ are the mean longitude of the object and the planet respectively. TP orbits librate around $\pm 60^{\circ}$, but for eccentric TP orbits their libration centers are displaced with respect to the equilateral locations at $\pm 60^{\circ}$ (Namouni and Murray 2000). HS orbits librate around $180^{\circ}$ whereas QS librate around $0^{\circ}$.

In the planar case, the families of TP, HS and QS orbits are disjoint, however, in a case of sufficiently inclined and eccentric orbits there can exist compound orbits which are unions of the HS (or TP) and QS orbits, and transient co-orbital orbits for which transitions between different families occur (Namouni 
1999, Namouni et al. 1999, Christou 2000, Brasser et al. 2004a, Wajer 
2009, 2010). This type of resonant behavior has been observed in the motion of several known near-Earth asteroids.

We searched numerically for objects which are, or will be in the near future (in the next 100 years), co-orbitals of Jupiter and experience transitions between different types of co-orbital motion. We have selected these objects which have astrometric data collected over a longer period than 1 year and we have identified among them seven asteroids and one comet. Additionally, we decided to investigate three more objects of a shorter interval of observations (asteroid \aee\ and two comets: \ar\ and \wc\ which were previously investigated by Kinoshita and Nakai (2007).

\section{Selection of objects and the method of numerical integration}
The aim of this study was to find and analyze transient, compound co-orbitals or quasi-satellites of Jupiter.
Therefore, we looked for objects located in the co-orbital area
associated with this planet, i.e. those objects which orbital
semimajor axis, $a$, satisfies the following condition:

\begin{equation}\label{crit1}
 -\epsilon\leq\Delta a = a_J - a \leq +\epsilon,
\end{equation}

\noindent where $\epsilon \simeq 0.35528$\,a.u. is the Jupiter's Hill
radius and $a_J$ -- semi-major axis of Jupiter orbit.

\noindent According to this criterion we primarily selected 3\,160
asteroids discovered by October 2010, including 1\,213 numbered and
1\,947 unnumbered objects, respectively taken from the AstDys
pages\footnote{http://hamilton.dm.unipi.it/astdys/}, and 24 comets
from the latest version of the "Catalogue of Cometary Orbits"
(Marsden and Williams 2008), 11 observed only in one apparition, and 13
numbered short-period comets. To exclude an abundant population of
Trojan asteroids from a pre-selected sample, we looked for objects
that during the next thousand years (in the period between 2010 --
3010) do not librate all the time around one of the triangular
Lagrange points, L$_4 $ or L$_5 $. In other words, they are not
Trojans all the time, but at least temporary move in HS, TP or QS within
the next 100 years, where the transitions are possible between all
these types of orbits as well as compound orbits are included. Of
the 3\,160 pre-selected objects we have found seven new 
potentially interesting objects: \cu, \sa, \ql, \gh, (32511)
2001 NX$_{17}$, 2006 SV$_{301}$ and \Larsen, as well as six
previously identified objects: \hr, \ug\, \qq, \aee, \wc\, and \ar. The
dynamical behavior of seven objects is analyzed in detail in this
paper. The reason why we decided not to include the two objects 2001 NX$_{17}$ and 2006 SV$_{301}$
is following. The first of them, 2001 NX$_{17}$, is not captured into 1:1 mean motion resonance with Jupiter up to
 2700, however, after this time about 11\% of VOs start to librate around L$_5$ point. The
second, 2006 SV$_{301}$, will be librate around L$_5$ for at least
1\,000 years and probably leave this state after the year 3000 (about 80\% of VOs show this behavior).

In addition, we expanded the asteroid search to objects whose
semi-major axes satisfy the condition $ \epsilon < |\Delta a| \leq 2
\epsilon $, since co-orbital objects may have temporarily increased
value of $|\Delta a|$, as in the case of 2003 YN$_{107}$ which was
temporarily captured into the 1:1 resonance with the Earth
(Connors et al. 2004). We have found 10 such asteroids (three
numbered: 944, 6144 and 145485, and seven unnumbered objects: 2000
EJ$_{37}$, 2005 TS$_{100}$, 2006 UG$_{185}$, 2008 UD$_{253}$, 2000
AU$_{242}$, 2002 AO$_{148}$, 2010 AN$_{39}$). None of these objects
is librating in the Jupiter co-orbital region at present, but \ug\
will be temporarily captured into 1:1 mean motion resonance with
Jupiter in the future. Thus, we have decided to include \ug\ to our
study.

It is well-known that the non-gravitational effects (hereafter NG
effects) play an important role in the motion of comets. They are
important in determining the osculating orbits from observations
even in a very short time scales, for short-period comets sometimes
involving only two consecutive apparitions. Therefore, we have
limited our analysis primarily to large perihelion distance comets
($q > 3$\,a.u.). Using Eq. 1 we selected sample of 24
comets where 13~have observational interval longer than 1~yr. In
this sample of 24~comets we had only four objects with $q > 3$\,a.u.:
186P/Garradd, \Larsen , 244P/Scotti ($=$2000 Y$_3$, 2010 Q$_1$) and
2004 FY$_{140}$. Of these four only \Larsen  ~moves in a dynamically
interesting trajectory. However, we decided to include to our
analysis also two comets discussed by Kinoshita and Nakai (2007), both
with significantly smaller perihelion distances than 3.0\,a.u. and shorter interval od data than 1~yr. Thus, we were able to
compare their results (based on a pure gravitational
calculations) with our own analysis based on observational material
from the same periods of time (the two comets were not observed
after 2005) and pure gravitational calculations (NG~effects for
these two comets are indeterminable, see below).

Finally, we have taken a detailed analysis of the eight asteroids
and three comets, where two of comets are only for comparison
with Kinoshita and Nakai (2007). For each object from this sample we
determined its osculating orbit (hereafter nominal orbit) based on
the available observational material taken from IAU Minor Planet
Center\footnote{http://www.minorplanetcenter.org/iau/
ECS/MPCOBS/MPCOBS.html}. To derive the nominal orbit we used
the least square orbit determination method where the equation of
object's motion have been integrated numerically using the recurrent
power series method (Sitarski 1989, 2002). In the case of comets
C/2002~AR$_2$, C/2003~WC$_3$ and \Larsen ~we try to determine the
NG~effects using the standard-model expressions for outgassing
acceleration acting on the comet in the radial (Sun-Comet)
direction, transverse and normal direction (for more details see
Marsden et al. 1973 and Yeomans et al. 2004).

Orbital elements (nominal orbit), together with the characteristics
of observational material used for their determination are given in
Table 1.

\begin{table}
\tiny \caption{\label{tab:elements}Orbital osculating elements of investigated objects.}
\renewcommand{\tabcolsep}{0.9mm}
\begin{centering}
\begin{tabular}{cccrrrrrr}
\hline
 Object         & observational            & q          & a                & e                & i                & $\Omega$         & $\omega$         & M      \tabularnewline
                & interval,                & [a.u.]       & [a.u.]             &                  &[deg]             &[deg]             & [deg]            &[deg]   \tabularnewline
                & Number of obs.,          &            &                  &                  &                  &                  &                  &        \tabularnewline
                &  RMS [arc sec]           &            &                  &                  &                  &                  &                  &        \tabularnewline
 {[}1]          &{[}2]                     &{[}3]       &{[}4]             & {[}5]            & {[}6]            & {[}7]            & {[}8]            &{[}9]   \tabularnewline
\hline
(241944)        & 1998 08 30 - 2010 09 16  & 3.59       &      5.228284816 &      0.312650873 &       32.9023865 &      315.0062251 &       60.8624810 &      338.3608103  \tabularnewline
2002 CU$_{147}$ & 130, 0\farcs 47          &            & $\pm$0.000000263 & $\pm$0.000000175 & $\pm$  0.0000227 & $\pm$  0.0000205 & $\pm$  0.0000484 & $\pm$  0.0000316  \tabularnewline
\hline
2007 GH$_{6}$   & 2006 01 06 -- 2007 05 11 & 2.85       &      5.300363874 &      0.462179871 &       25.5220714 &       79.1745253 &       97.5704892 &      109.1506872  \tabularnewline
                &  34, 0\farcs 78          &            & $\pm$0.000350482 & $\pm$0.000031212 & $\pm$  0.0002031 & $\pm$  0.0003596 & $\pm$  0.0022926 & $\pm$  0.0103139  \tabularnewline
\hline
2006 QL$_{39} $ & 2006 07 21 -- 2010 05 05 & 2.04       &      5.119417841 &      0.600803824 &       13.3515816 &      172.5101475 &      253.9149895 &      106.3479874  \tabularnewline
                & 524, 0\farcs 34          &            & $\pm$0.000008572 & $\pm$0.000000650 & $\pm$  0.0000106 & $\pm$  0.0000199 & $\pm$  0.0000399 & $\pm$  0.0002740  \tabularnewline
\hline
2006 SA$_{387}$ & 2000 02 06 -- 2006 10 29 & 4.05       &      5.028032303 &      0.194170169 &        3.8408196 &      130.7342828 &      199.1410830 &      149.0652766  \tabularnewline
                &  29, 0\farcs 17          &            & $\pm$0.000051035 & $\pm$0.000009109 & $\pm$  0.0000163 & $\pm$  0.0002665 & $\pm$  0.0037205 & $\pm$  0.0002764  \tabularnewline
\hline
2001 QQ$_{199}$ & 2001 08 22 -- 2009 05 30 & 3.03       &      5.328276992 &      0.430755163 &       42.4807691 &      213.0903290 &      192.8582380 &      249.1814160  \tabularnewline
                & 144, 0\farcs 51          &            & $\pm$0.000005209 & $\pm$0.000000406 & $\pm$  0.0000367 & $\pm$  0.0000151 & $\pm$  0.0001442 & $\pm$  0.0004217  \tabularnewline
\hline
2004 AE$_{9}$   & 2003 12 18 -- 2004 03 16 & 1.82       &      5.109087756 &      0.644348546 &       1.6494590   &      188.7027920 &      285.7809178 &      204.8711444  \tabularnewline
                & 81, 0\farcs 28           &            & $\pm$0.000042932 & $\pm$0.000013097 & $\pm$  0.0000598 & $\pm$  0.0024734 & $\pm$  0.0103257 & $\pm$  0.0147365  \tabularnewline
\hline
(118624)        & 1960 09 24 -- 2010 04 15 & 4.09       &      4.958976214 &      0.174840974 &       15.5203750 &      223.4163388 &      353.9220196 &      334.2833720  \tabularnewline
2000 HR$_{24}$  & 168, 0\farcs 51          &            & $\pm$0.000000200 & $\pm$0.000000191 & $\pm$  0.0000209 & $\pm$  0.0000367 & $\pm$  0.0000472 & $\pm$  0.0000356  \tabularnewline
\hline
2006 UG$_{185}$ & 2002 06 29 -- 2008 03 13 & 4.21       &      4.828518506 &      0.127187519 &       20.0199173 &      131.7842702 &      301.2250799 &      168.9330187  \tabularnewline
                &  44, 0\farcs 36          &            & $\pm$0.000032545 & $\pm$0.000008592 & $\pm$  0.0000643 & $\pm$  0.0000457 & $\pm$  0.0011556 & $\pm$  0.0029683  \tabularnewline
\hline\hline
\Larsen & 1997 11 03 -- 2009 01 21 & 3.27       &4.912958883       &      0.333647494 &       12.1162513 & 234.8063310      &      133.8893053 &       62.9989331  \tabularnewline
          & 552, 0\farcs 64          &            & $\pm$0.000002874 & $\pm$0.000001298 & $\pm$  0.0000482 & $\pm$ 0.0001643  & $\pm$  0.0018584 & $\pm$  0.0009011  \tabularnewline
\hline
P/2002 AR$_{2}$ & 2002 01 06 -- 2002 04 07 & 2.06       & 5.346332847      & 0.615331257      &       21.0953152 &        7.7046084 &       73.2460548 & 248.9045388  \tabularnewline
LINEAR          &  42, 0\farcs 55          &            & $\pm$0.001971033 & $\pm$0.000131378 & $\pm$ 0.0002818  & $\pm$  0.0020963 & $\pm$ 0.1400495  & $\pm$  0.1400495 \tabularnewline
\hline
P/2003 WC$_{7}$ & 2003 11 18 -- 2004 04 11 & 1.67       & 5.195904282      & 0.678999002      &       21.4262026 & 88.8007066       & 342.2159165      &      196.7368846  \tabularnewline
LINEAR-Catalina & 108, 0\farcs 67          &            & $\pm$0.000471413 & $\pm$0.000028464 & $\pm$  0.0000658 & $\pm$ 0.0001072  & $\pm$ 0.0003677  & $\pm$ 0.0269043  \tabularnewline
\hline
\end{tabular}
\end{centering}
Overall description of the
observational material (column [2]) and osculating orbital elements
of investigated objects (nominal orbits, columns [4]--[9]) at
the epoch of 2010 07 23.0 TT$=$ JD 2455400.5 where the standard
notation for asteroids was applied: $a$ denotes semimajor axis, $e$
-- eccentricity, $i$ -- orbit inclination, $\Omega$ -- longitude of
the ascending node, $\omega$ -- argument of perihelion and $M$ --
mean anomaly; in the column [3] the perihelion distance is
additionally given. Angular orbital elements are referred to the
equinox J2000.0. In the second line of column [2] the number of
observation and root-mean-square of the nominal orbit fit to the
data (RMS) are presented for each object
\end{table}

An analysis of the uncertainties of orbital elements given in
Table 1 shows that four asteroids (two
numbered, \cu, \hr, and two unnumbered, \qq , \ql) and comet
\Larsen\ have orbits very well determined; the worst determined
is the orbit of comet C/2002~AR$_2$ taken only to compare with the
result of Kinoshita and Nakai (2007).

In the case of comets, we realize that the only osculating orbit of
the comet 200P/Larsen is based on the data taken from a
sufficiently long time interval to draw conclusions about its
dynamical evolution both because of their orbital uncertainties
as well as large perihelion distance, which suggests that the
NG~effects less affect pure gravitational motion of this comet than
in the case of the remaining two comets.

Comets C/2002~AR$_2$ and C/2003~WC$_7$ have perihelia at 2.1\,a.u. and
1.7\,a.u. from the Sun, respectively. Thus, the NG~effects may be
significant in the motion of these comets. Unfortunately, the
 short time intervals of astrometric observations (less
than half a year) do not allow us to determine their
NG~orbits. Thus, these two comets have placed in the Table
1 after Kinoshita and Nakai (2007). only as a potentially
interesting objects.

\noindent Although the comet 200P/Larsen has a perihelion farther
from the Sun than those two (at 3.3\,a.u. from the Sun), the
NG~effects are determinable by a much longer time interval of
observations (more than 11 years). However, we calculated that
taking into account the NG~effects does not cause a significant
decrease of RMS (only $\sim0\farcs 0$4; from $0\farcs 64$ to
$0\farcs 60$). Nevertheless, comparing the pure gravitational
dynamical evolution starting from the NG~osculating orbit with
that starting from the purely gravitational osculating orbit
it seems obvious that we know very well the dynamical evolution of
this comet only 94 years back (after the comet encounter with
Jupiter in 1917) and for 230 years forward (to Jupiter close
encounter in the year 2242).
Therefore, the detailed discussion of the dynamical evolution of
this comet (section 3.2) is constrained just to that period.

To examine the past and future dynamical evolution of investigated
objects two hundred of virtual object (VO)
orbits using the Sitarski's method of the random orbit selection
(Sitarski, 1998) were examined. The derived swarm of
fictitious objects (VOs) follows the normal distribution in the
space of orbital elements. Also the RMS's fulfil the 6-dimensional
normal statistics (thus, all VOs are compatible with observations;
for more details see also Kr\'olikowska et al. 2009). Thus, as the starting data
for our dynamical calculations we constructed (for each investigated
object) osculating swarms of 201~VO orbits (including nominal orbit
presented in Table 1).

The numerical orbit computations of the individual VO motion
were performed using the recurrent power series (RPS) method
(Hadjifotinou and Gousidou-Koutita 1998) by taking the perturbations by all eight planets,
the Moon and Pluto\footnote{Pluto was included to obtain
compatibility with the ephemeris DE406.} into account (similarly as was taken for the osculating orbit determination). The RPS algorithm allows to determine the optimal value of the integration steps at each point along the orbit, ensuring the
desired accuracy of results of computations is obtained. This method
also give very good results in the case of close approaches of an
object to the Sun and the planets (Sitarski 1979).

The objects and their VOs were treated as massless test particles.
Since we focused mainly on the future dynamical evolution of the
objects, the maximum integration length was 2000 Julian years
backward only and 10000 years forward. The time interval of our
integrations depends on the predictability time (i.e., qualitatively speaking, within the period where all VOs show very similar behavior) and was selected from a practical point of view (see section 3 as well as table 2 for the integration time).

\section{Analysis of the results}
\label{analysis}
\subsection{Transient quasi-satellite-tadpole types of orbits: asteroids \cu\ and \gh}
\label{qs_tp}
The asteroid \cu\ is currently not classified but the value of its Tisserand parameter with respect to Jupiter, T$_J=2.60$\footnote{which is defined by $T_J=a/a_j+2\sqrt{(a/a_J)(1-e^2)}\cos I $, where $a$ and $a_J$ are the semimajor axis of the object and Jupiter, respectively, $e$ is the eccentricity, and $I$ is inclination relative to the orbital plane of Jupiter of the object's. At the moment of proof correction this asteroid is classified as Main Belt Asteroid as the IAU Minor Planet Center Web page.} suggests that it is asteroids in cometary orbits (ACO), see Sect. 4 for definition and discussions. This asteroid is currently a QS of Jupiter for about 200 years and will be in this state up to 2030 (Fig. 1, left panels). In this state the object is moving deep inside the co-orbital region ($|\Delta a|<0.1\epsilon$) with amplitude of the principal resonant angle $\sigma\simeq4^{\circ}$. Near 2030, the transition from QS to TP (libration around L$_4$ point) motion occurs. This TP phase lasts about 350 years and after 2650 the asteroid switches to librate around L$_5$ Lagrange point.

Figure 1 (left figures, top and middle panels) shows that as far as the past and future dynamical evolution of \cu\ is concerned, i.e. within the time interval [1000-7000], its motion undergoes transitions between two types of trajectories, QS and TP regimes, where (in TP type of motion) the guiding center\footnote{The motion of an asteroid in a co-orbital region is a superposition of two components: a short period three-dimensional
epicyclic motion with amplitudes of $\sim ae$ and $\sim a \sin i$, and a long period
term corresponding to the averaged position of the asteroid with respect to the planet known as the
guiding center of the motion. The motion of the guiding center is represented by the variables $\Delta a$ and $\sigma$ (Namouni et al. 1999, Christou 2000).} of the asteroid librates either around L$_4$ or L$_5$ Lagrangian points. In the QS phases the asteroid's change of semimajor axis is of order $|\Delta a|<0.1\epsilon$, and the amplitude of libration of $\sigma$ is smaller than $12^{\circ}$. On the other hand, in the TP regimes, one has $|\Delta a|<0.5\epsilon$ and the amplitude of libration of $\sigma$ is about $50^{\circ}$. At present the eccentricity and the inclination of the \cus\ is about 0.31 and 33$^{\circ}$, respectively. The eccentricity changes between 0.11 and 0.37, the inclination - from about 30$^{\circ}$ to 36$^{\circ}$.

\begin{figure}[htbp]
\centerline{\includegraphics [width=14cm]{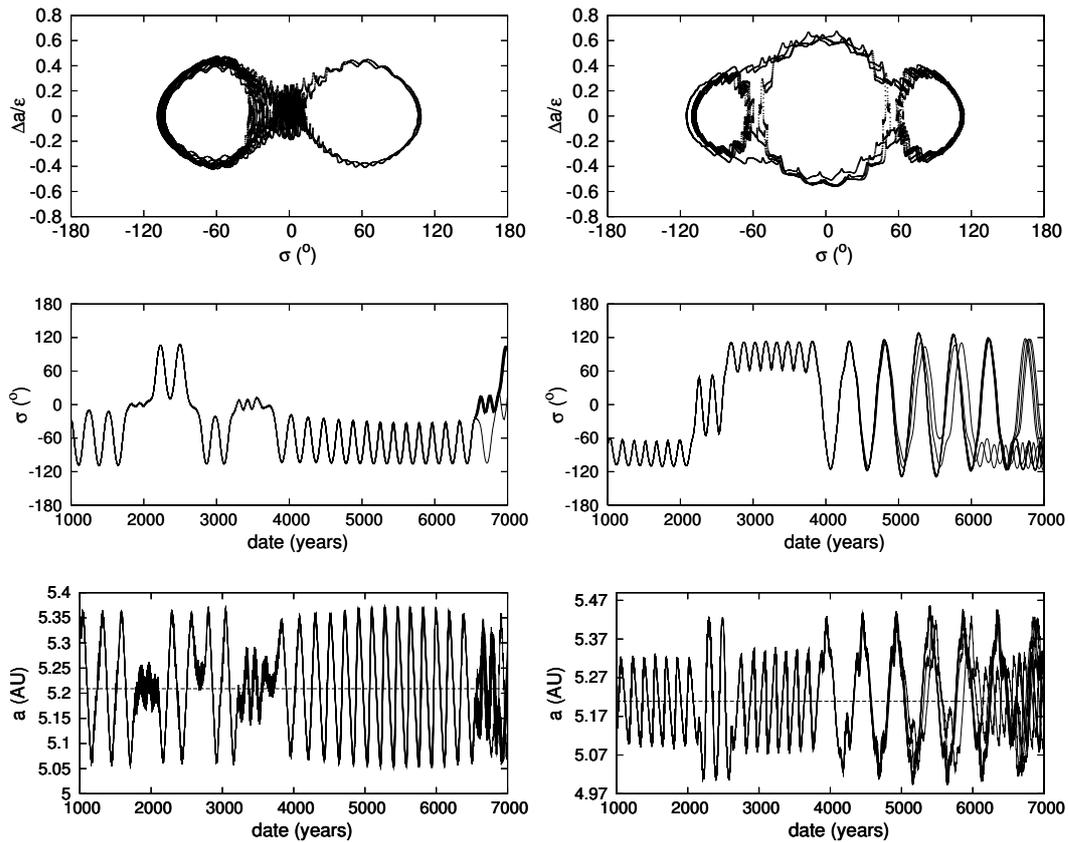}}
\caption{Asteroids \cu\ (left) and \gh\ (right). Top: evolution of the guiding
center of the asteroid. The time interval is the same as in the case of both middle and bottom panels, middle: evolution of the principal resonant angle of the representative subsample of 10 VOs (nominal
orbit plus 9 randomly selected VOs), bottom: evolution of the
semimajor axis of 10 VOs. The dashed line indicates the semimajor
axis of Jupiter.}
\label{Fig1}
\end{figure}

Our calculations show that all VOs of \cu\ move on similar orbit up to about 6500, after that time they still move inside the co-orbital region of Jupiter but they show different behavior (Fig. 1 left figures, middle and bottom panels). We obtained that $14\%$ of the VOs will continue the motion around L$_5$ point (starting from the year 3750), the remaining $86\%$ will transit into orbits which librate around L$_4$ point. During the predictability time interval, the asteroid does not experience close approaches with Mars, Jupiter and Saturn. The smallest distances between each of these planets and \cu\ are over 1.80~a.u., 1.25~a.u. and 2.18~a.u., respectively.

The asteroid \gh\ is currently moving on a high eccentric and moderate inclined orbit (e=0.46, i=25.52$^{\circ}$, $T_J=2.63$) and, for the previous 1000 years at least, it has been librating on a TP orbit in the L$_5$ area. As is illustrated in Fig. 1 (right figures, top and middle panels) near 2150, it transits from TP to QS regime. After completion 2.5 QS loops, about the year 2600, the object starts to librate in TP orbit around the L$_4$ lagrangian point. The amplitude of libration is 25$^{\circ}$ when \gh\ is in the TP phase and 50$^{\circ}$ in the QS phase. Near 3900 this asteroid leaves QS state and will be moving in TP-QS-TP orbit with large amplitude of $\sigma$.

Almost 1.5 year interval of observations of this asteroid is relatively short but all VOs present a very similar behavior up to 4500 year (see Fig 1, right figures, middle and bottom panels). After that moment all VOs of \gh\ start to diverge, nevertheless, about 50$\%$ of them still continue the motion in the TP-QS-TP orbit and about 4$\%$ is locked in the $L_5$ area until the end of the integration.

\subsection{Temporary horseshoe types of orbits: asteroids \ql, \sa\ and comet 200P/Larsen}
\label{qs_hs}
These three objects, two asteroids and one comet, are the only known co-orbitals of Jupiter which are currently trapped in a HS orbit. As one can see in top and bottom panels of Fig. 2 and  Fig. 3 their semimajor axis alternates between slightly smaller and$\backslash$or slightly larger values than the outer boundaries of the co-orbital region (4.85 a.u. - 5.56 a.u.). In general, these boundaries separate HS orbits (or another types of co-orbital motion) from (irregular) passing ones. Thus, these boundaries can be interpret as the stability limits of the co-orbital motion (see also Namouni et al. (1999), Connors et al. (2004)).

\ql\ has very large value of eccentricity ($e=0.60$), small value of inclination ($i=13.35^{\circ}$) and the Tisserand parameter with respect to Jupiter, $T_J$=2.53. At present it is in a compound HS-QS orbit as one can see in Fig. 2 (left figures, top and middle panels). The object transited from the QS to the HS state about 1920. It stays in the HS orbit to 2330 and then the object enters a QS orbit. The asteroid has interval of observations of about 4 years. Within the predictability period (see below) the asteroid experiences a few close encounters ($<0.6$ a.u.) with Jupiter. The close approach to Jupiter took place about 1600 ($\sim0.4$ a.u.) and other encounters will be about 2630 and 2791 when the asteroid will pass the planet at a distance of 0.32 a.u. and 0.22 a.u., respectively. This object also experiences, because of large value of its eccentricity, close approaches with both Mars ($\sim$ 0.6 a.u.) and Saturn ($\sim$2.2 a.u.). We obtained that after the close approach of \ql\ to Jupiter about the year 2790 the orbital elements start to diverge significantly and the orbit of this object becomes unpredictable. Going back in time the orbital evolution of \ql\ is predictable for about 700 years, i.e. back to the year 1300.

\sa\ has moderate eccentricity  ($e = 0.19$), small inclination ($i = 3.84^{\circ}$) and the Tisserand parameter
with respect to Jupiter, $T_J$=2.89. Its observational interval is about 6 years. The nominal orbit and all VOs show a very similar behavior within the time interval of [1780, 2250], while outside this period dynamical evolution of all VOs starts to diverge (see Fig. 2 right figure, middle and bottom panels). Within this almost five hundred year period the object executes two horseshoe loops, the first slightly outside the co-orbital region ($|\Delta a|<1.10\epsilon$) and the second inside ($|\Delta a|<0.95\epsilon$)(Fig. 2 right figures, top panel) as well as experiences several moderately deep ( $< 0.6$ a.u.) encounters with Jupiter. Forward in time about 50$\%$ of VOs is moving on the horseshoe orbit to 2500. Then, almost all VOs leave the co-orbital region of Jupiter, however about $7\%$ of VOs stay inside the co-orbital region to 3500 and transit between HS and TP types of motion.

\Larsen\ belongs to the Jupiter family comets. It has moderate eccentricity ($e=0.33$) as well as inclination ($i=12.11^{\circ}$), and the Tisserand parameter with respect to Jupiter is $T_J$=2.74. Fig. 3 shows that before the year 1995 this comet was moving outside Jupiter co-orbital region. In 1995 it passed at the distance of 0.35 a.u. from the giant planet. This very close encounter with Jupiter caused the shortening of the semimajor axis of the comet orbit from 5.7 a.u. to 4.9 a.u. and a decrease in the perihelion distance from 4.0 a.u. to 3.3 a.u. Finally, the comet has started to move in an HS orbit. After its HS episode, lasting about 135 years, the transition into QS orbit will take place (around the year 2130) from the L$_5$ lagrangian point. \Larsen\ executes two QS loops and leaves this state near 2300 and again transits into a HS regime.

Because of the moderate eccentricity \Larsen\ do not pass very close both Mars and Saturn. The minimal distances between the comet and these planets are over 2.0 a.u. and 3.0 a.u., respectively. On the other hand, the object experiences multiple close encounters with Jupiter. The closest approaches with this planet occurred in 1917 (0.063 a.u.) and in 1995 (0.35 a.u.). The former encounter changed the semimajor axis from 8.8 AU to 5.8 AU and perihelion distance from 4.94 a.u. to 4.03 a.u.. When this comet will stay in a QS orbit (2130 -2300) the distance of \Larsen\ from Jupiter will remain larger than $\sim0.7$ a.u.

According to presented analysis the orbit of \Larsen\ seems to be
predictable within the time interval of [1917-2800] (Fig. 3, bottom
panel). However, comparing pure gravitational evolutionary
tracks for two starting nominal orbits, NG and pure
gravitational, we conclude that the orbit of this comet is known well only 100
years back (after the comet encounter with Jupiter in 1917) and 230 years forward (to Jupiter close encounter in the year 2242) (see also discussion in Sect.2).

\begin{figure}[htbp]
\centerline{\includegraphics [width=14cm]{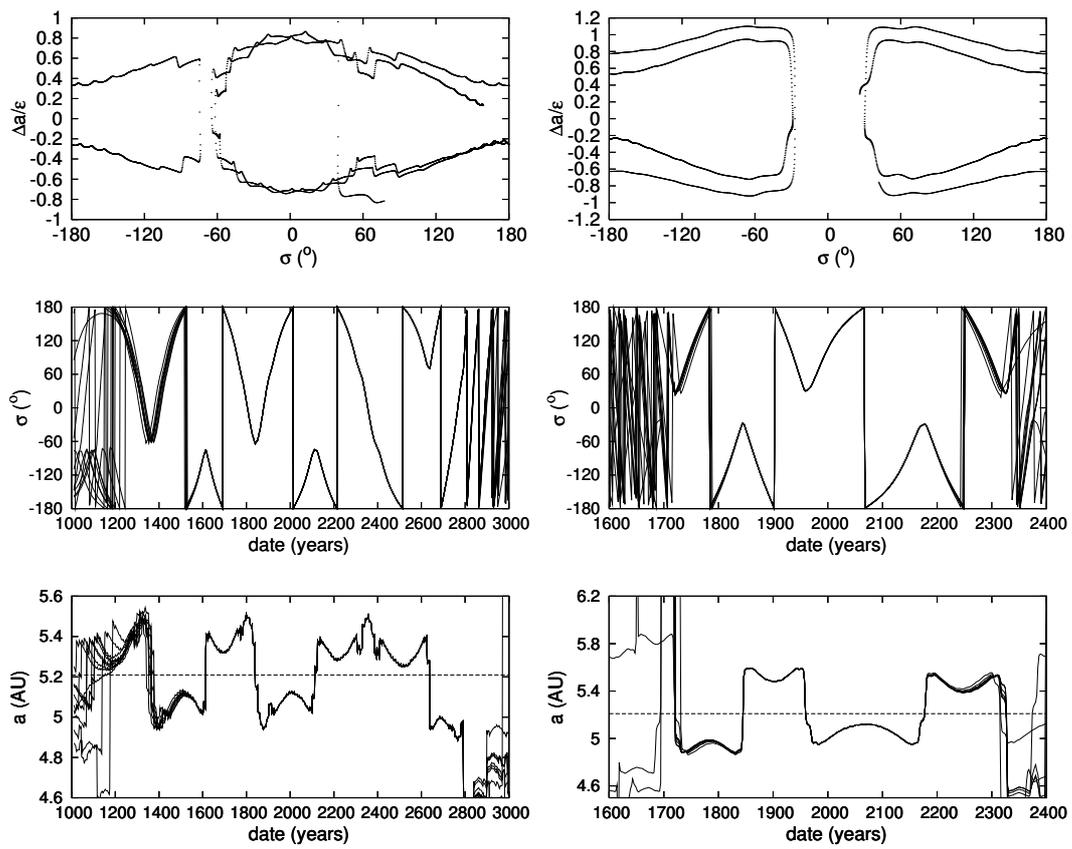}}
\caption{The same as in Fig.~1 for asteroids \ql\ (left) and \sa\ (right).}
 \label{Fig2}
\end{figure}

\newpage

\begin{figure}[htbp]
\centerline{\includegraphics [width=7cm]{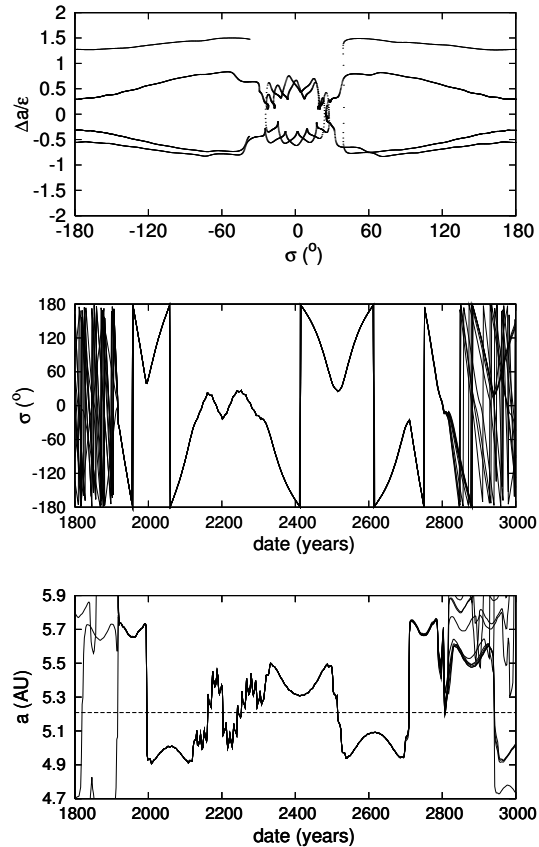}}
\caption{The same as in Fig.~1 for comet \Larsen.}
\label{Fig3}
\end{figure}

\subsection{Present quasi-satellite objects: asteroids \qq, \aee\ and comets \ar,  \wc}
\label{qs}

\qq\ is currently a quasi-satellite of Jupiter (Kinoshita and Nakai 
2007). Its amplitude of $\sigma$ varies from $105^{\circ}$ to $115^{\circ}$ with libration period of about 490 years (see Fig. 4 left figures, top and middle panels). The orbital eccentricity of the asteroid is $e=0.43$, the inclination $i=42.48^{\circ}$ and the Tisserand parameter, $T_J=2.37$. The minimum distance of the asteroid from Jupiter is larger than $\sim$1.34~a.u. and from Mars and Saturn larger than $1.0$ a.u. Kinoshita and Nakai (2007) found that around October 2013 \qq\ approaches to the Earth at a distance of 2.4 a.u. and they expect to detect a cometary activity of this object.

Relatively long interval of observations of about 8 years of this object and lack of deep close approaches to planets allows to determine the dynamical behavior for a long period of time. As one can see in Fig. 4 (left figures, middle and bottom panels) the nominal orbit and all VOs present a very similar behavior within the assumed time of integration i.e [0-12000] but at the end of integration the VOs start to slightly diverge.

\aee\ is the second example of asteroids to be recognized as a QS of Jupiter by Kinoshita and Nakai (2007). As one can see in Table 1 this asteroid has very high eccentricity ($e=0.64$) and very small inclination ($i=1.65^{\circ}$). Its Tisserand parameter with respect to Jupiter is about 2.50. Within assumed time of integration (i.e. in years 0-12000) its eccentricity varies from 0.55 to 0.72 and the inclination decreased from $10^{\circ}$ to the present value. In the future the inclination will increase up to $16^{\circ}$. This object is not quite a Mars crosser, but it experiences very close encounters ($\sim$0.1 a.u.) with this planet. The next close approach to Mars will occur in 2015 at a distance of 0.57 a.u. The distance of \aee\ from Jupiter and Saturn remains larger than $\sim$1.6 a.u. and $\sim$0.5 a.u., respectively. Kinoshita and Nakai (2007) found that in July 2015 this object will approach to the Earth within 1.81 a.u. and they predict that a cometary activity of this object can be detected.

In the QS regime the asteroid librates with amplitude of principal resonant angle $\sigma = 35^{\circ}-60^{\circ}$ and a period of libration of about 200 years. During every QS loop its semimajor axis alternates regularly from 4.99 to 5.43 a.u. as one can see in Fig. 4 (right figures).

According to our numerical simulations the nominal orbit and all of VOs present a very similar behavior over the entire range of integration [0, 12000]. Kinoshita and Nakai (2007) showed that \aee\ will occupy its current QS orbit to 15000.

Comets \ar\ and \wc\ were found as asteroids and then a cometary activity was detected on these bodies. Both comets belong to Jupiter family comets with the Tisserand parameter of 2.52 and 2.36, respectively. These objects are currently Jupiter QS (Kinoshita and Nakai 2007). However, both comets have short interval of observations (see Table 1) what means that it is not possible to determine the NG effects for these objects.
Their perihelion distances are 1.7 \,a.u. and 2.1 \,a.u. from the Sun, respectively. The evolution of the guiding center of both comets, the changes in semimajor axis and the principal resonant angle of 10 randomly selected VOs are presented in Fig. 5 within a time interval of 1000 years backward and forward. However, we must keep in mind that due to the calculated uncertainties of orbital elements, the NG~effects, which can be significant in the motion of these comets, and due to the interactions with the planets, it is clear that we cannot trace the orbits of these objects with certainty over a period larger than a few hundred years.

\ars\ is moving on a regular QS orbit with amplitude of the guiding center of about $50^{\circ}$ deep inside co-orbital region ($|\Delta a|<0.6\epsilon$) (Fig. 5, left figures). This comet does not experience close approaches to both Jupiter (the distance of this object from the Jupiter is larger than $2.0$ a.u. during the integration period) and Saturn ($>$3.0 a.u.) but it makes multiple relatively close approaches ($>$0.6 a.u.) to Mars. In 2014, the comet reaches its the closest approach distance of 1.9 a.u. to the Earth.

Comet \wcs\ is librating in a QS orbit (right middle panel of Fig. 5). Near 2500, the orbital elements of this comet start to diverge and a subsequent transitions into TP orbits are possible (Fig. 5, right figure, top and middle panels). Our calculations indicate that in the past and the next a few hundred years its distance from Jupiter, Saturn and Mars is never less than 2.0 a.u., 3.0 a.u. and $\sim$0.6 a.u., respectively. The comet will make its closest approach to the Earth (0.71 a.u.) in the year 2015.

\begin{figure}[htbp]
\centerline{\includegraphics [width=14cm]{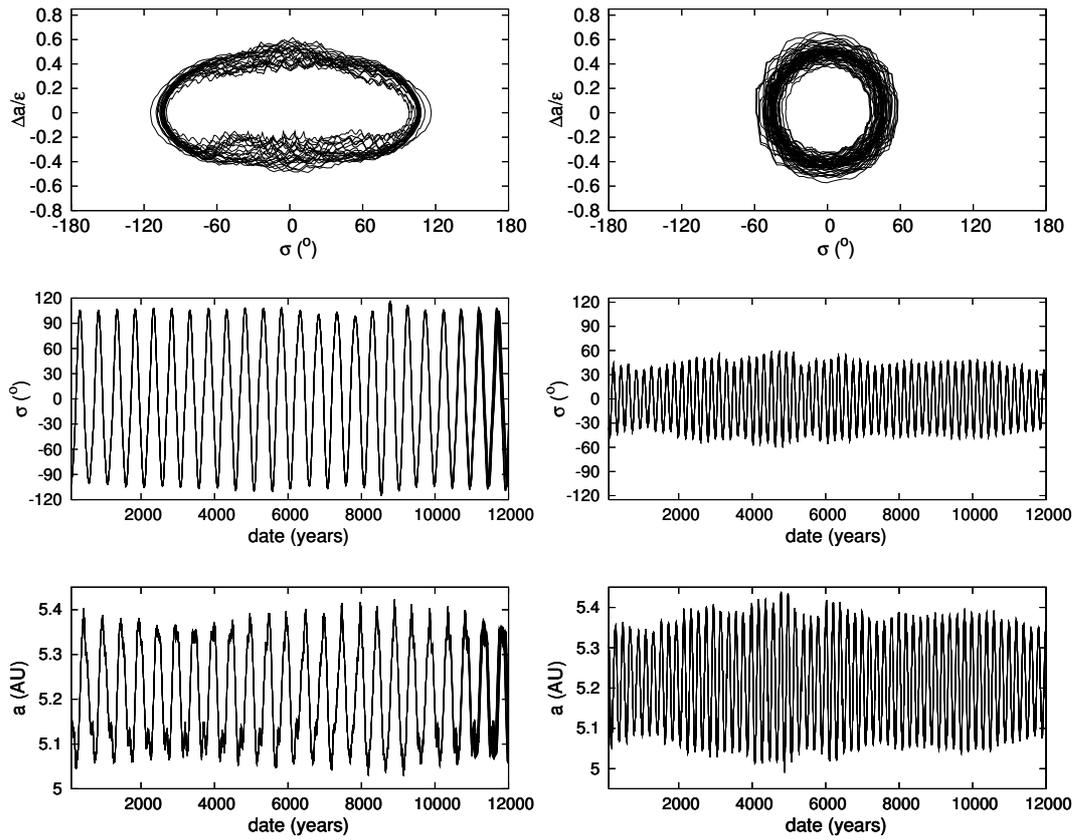}}
\caption{The same as in Fig.~1 for asteroids \qq\ (left) and \aee\
(right).} \label{Fig4}
\end{figure}

\newpage

\begin{figure}[htbp]
\centerline{\includegraphics [width=14cm]{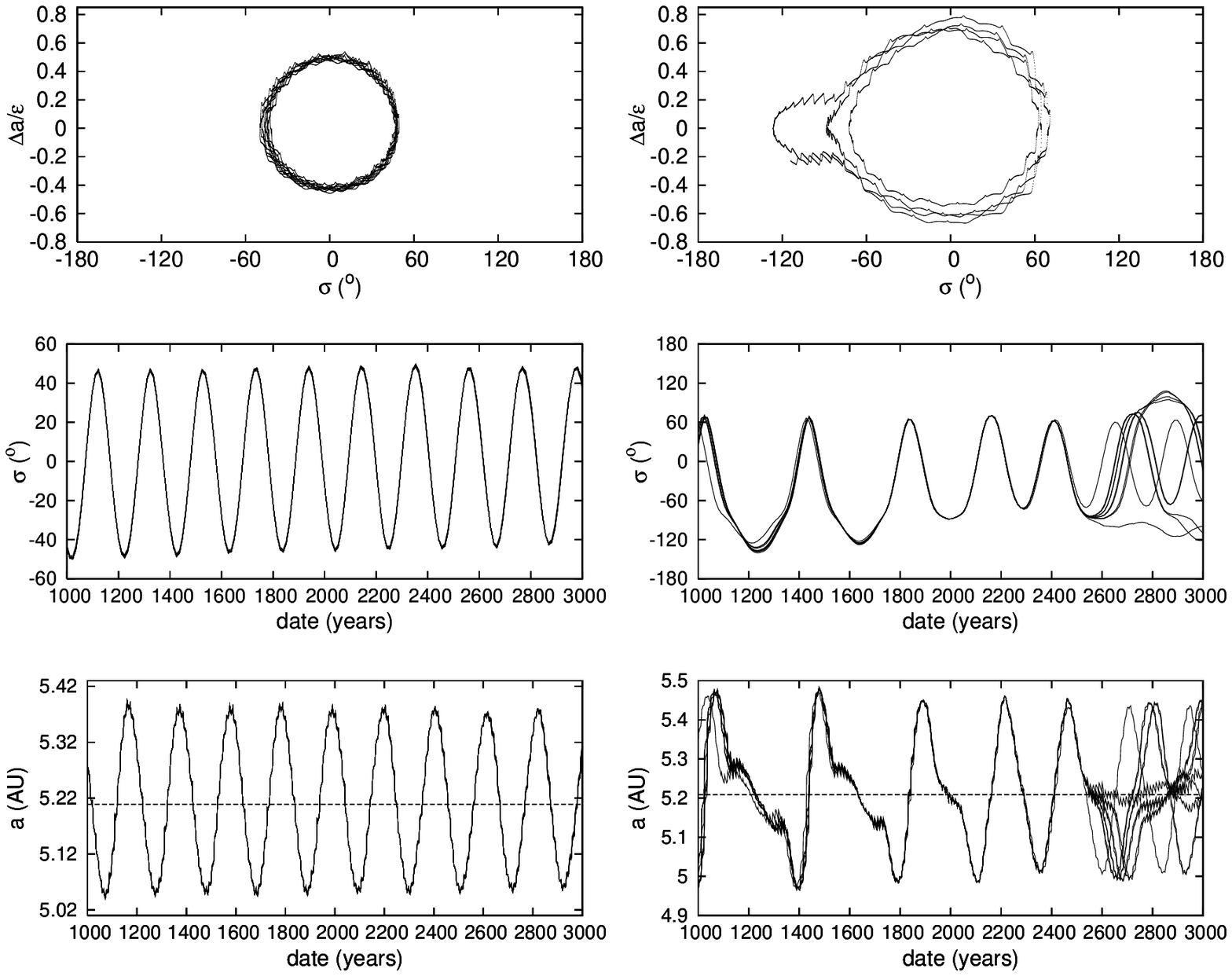}}
\caption{The same as in Fig.~1 for comets \ar\ (left) and \wc\
(right).} \label{Fig5}
\end{figure}

\subsection{Temporary captured co-orbital objects: \hr\ and \ug}
\label{temporary_qs}
Currently, the asteroids \hr\ and \ug\ are moving in passing-type orbits (see both middle and bottom panels of Fig. 6). They have moderate values of eccentricity and inclination (see Tab. 1) and the Tisserand parameter with respect to Jupiter is 2.80 and 2.72, respectively.

The interval of observations of \hr\ is about 50 years, thus its orbit is well known. However, due to close multiple encounters with Jupiter the swarm of VOs starts to diverge after the year 2350 and before the year 1350 as one can see in Fig. 6 (left figures, middle and bottom panels). The boundaries of the predictability time are comparable with these in Karlsson (2004). The semimajor axis of this object had been slowly increasing towards Jupiter's value of semimajor axis and finally this object entered Jupiter's co-orbital region near 1913 from the L$_5$ side (Fig. 6, left figures, middle and top panels). Near 2050 \hr\ will be temporarily trapped in 1:1 mean motion with Jupiter in a QS orbit as was also found by Karlsson (2004). The object leaves this state (as well as the co-orbital region of Jupiter) near 2136, completing one QS loop. In the QS state, \hr\ experiences multiple close approaches to Jupiter; the closest encounters with the planet will occur in 2095 (0.57 a.u.), 2107 (0.53 a.u.) and 2136 (0.22 a.u.).

Our calculations show that, in the near future (i.e. to 3000), about 58$\%$ of VOs have semimajor axis smaller (mostly between 4 and 5 a.u.) than that of Jupiter, in the past (i.e. in the years 1000-2000) - 71$\%$. This behavior is also in good agreement with predictions by Karlsson  (2004).

\begin{figure}[htbp]
\centerline{\includegraphics [width=14cm]{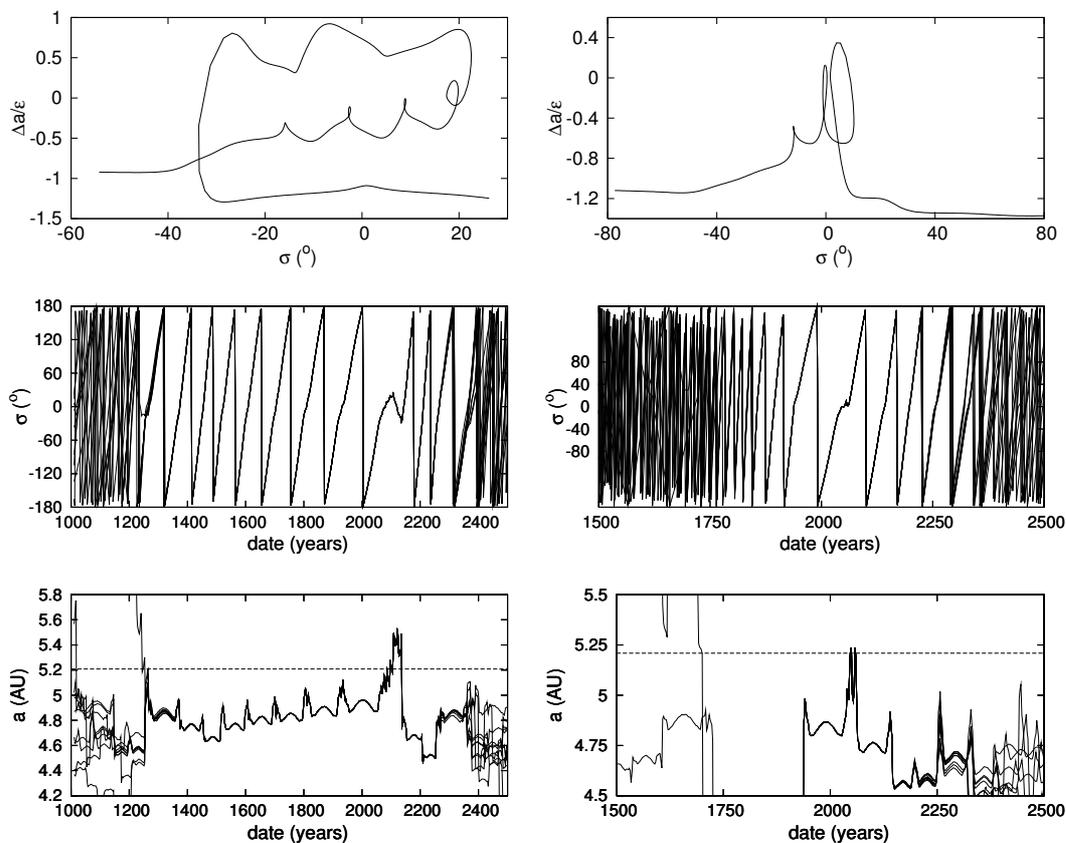}}
\caption{The same as in Fig.~1 for asteroids \hr\ (left) and \ug\
(right). In the case of \hr\ and \ug\ the time evolution of their guiding center is
plotted within the time interval 2050-2150 and 2020-2080 respectively.}
\label{Fig6}
\end{figure}

\ug\ is the second example of an object which will be temporarily captured by Jupiter into its co-orbital area in the next 40 years. In this case, we had a significantly shorter period of 6 yr observations than for the previous object, however, the period of predictability of \ug\ seems to be comparable with that for \hr. One can see in Fig. 6 (right figures, middle and bottom panels) that the dynamical evolution of the orbital elements of \ug\ can be predictable from 1800 to 2320. The transition into co-orbital orbit occurs around 2045 and lasts about 15 years. The co-orbital part of the motion is visible in Fig. 6 (right figures, top and middle panels) as a one revolution of the guiding center with small value of the principal resonant angle ($|\sigma|<1^{\circ}$). During the co-orbital episode \ug\ remains within the distance of 2.0 a.u. from Jupiter. In this state it makes closest approach to Jupiter near 2056, passing at a distance of about 0.31 a.u. from the planet. Although its guiding center enclosed origin (smaller left loop at Fig. 6, right figure, top panel), this object is not captured in a QS orbit. In the QS orbit an object should stay about 12 at least (orbital period of Jupiter) - the smaller left loop of \ug\ last about 6 years (2045-2051).

In the past, i.e. in the years 1000-2000, 81$\%$ of VOs had a semimajor axis smaller than that of Jupiter and did not experiences co-orbital episodes. In the future, after leaving the co-orbital temporary capture state, until the end of our integration the fraction of these objects drops to just 23$\%$. The remaining VOs may be captured into co-orbital area.

\section{Summary and discussion}
\label{summary}
We have analyzed the orbits of six already known, and we have identified five new non-Trojan, co-orbitals of Jupiter. These object have been divided into several classes according to their dynamical properties. We selected four such classes of the co-orbital behavior of the objects: TP$/$QS, HS$/$QS, long lasting QS and temporary co-orbitals. The most important results of our investigation are summarized in the Table 2. In this table are listed the Tisserand parameter with respect to Jupiter, the type of co-orbital motion, time of integration and the estimated period of predictability of the all considered objects.

The analyzed objects can also be classified according to their Tisserand parameter with respect to Jupiter, $T_J$. According to this criterion cometary orbits are defined as those having $T_J < 3$,  while asteroidal orbits are those with $T_J > 3$. Therefore, all the objects with $T_J < 3$ that do not present any signature of cometary activity are defined as an asteroids in cometary orbits (ACO) (Licandro et al. 2006). They also found that ACO with $T_J>2.9$ have spectra typical of the main belt objects while those with $T_J < 2.9$ shown comet-like spectra. All objects investigated here have $T_J < 2.9$, with the largest value of $T_J = 2.89$ for asteroid 2006 SA$_{387}$.

The three analyzed comets belong to the Jupiter family comets ($T_J>2.0$) and, as one can see in the Table 2, the Tisserand parameter of analyzed asteroids also locate all of them as Jupiter family comets. In fact, the comets \wc\ and \ar\ were discovered as asteroids 2003 WC$_7$ and 2002 AR$_2$, and later a cometary activity was recognized on these objects (Kinoshita and Nakai 2007). One may expects that the analyzed asteroids, in the future, also can exhibit cometary outgassing.

\begin{table*}
\tiny
\caption{\label{tab:dynamics}The Tisserand parameter, type of co-orbital behavior, integration and predictability period for the analyzed objects.}
\begin{tabular}{l*{4}{c}r}
Object              & T$_J$ & Dynamical  & Integration & \multicolumn{2}{c}{Predictability period}    \\
 &            & behavior &  period &  from & to  \\
\hline
(241944) 2002 CU$_{147}$ & 2.60 & transient TP-QS & 1000-7000  & $<$1000&6500   \\
\hline
2007 GH$_{6}$ & 2.63 & transient TP-QS, compound TP-QS-TP & 1000-7000 & $<$1000&5500\\
\hline
2006 QL$_{39}$ & 2.53 & temporary compound HS & 1000-3000 & 1200&2791 \\
\hline
2006 SA$_{387}$ & 2.89 & temporary HS & 1000-3000 & 1780&2250 \\
\hline
2001 QQ$_{199}$ & 2.37 & long-lasting QS & 0-12000  & $<$0&$>$12000 \\
\hline
2004 AE$_{9}$ & 2.50 & long-lasting QS & 0-12000 & $<$0&$>$12000 \\
\hline
(118624) 2000 HR$_{24}$ & 2.80 & temporary QS & 1000-3000 & 1350&2350   \\
\hline
2006 UG$_{185}$ & 2.72 & temporary co-orbital & 1000-3000 & 1800&2320 \\
\hline\hline
200P Larsen     & 2.74 &transient QS-HS & 1000-3000 & 1917&2800    \\
\hline
C/2002 AR$_{2}$ LINEAR     & 2.52 & long-lasting QS & 1000-3000 & $<$1000& $>$3000    \\
\hline
C/2003 WC$_{7}$ LINEAR-Catalina     & 2.36 & compound TP-QS$/$QS & 1000-3000 & 1300&2500 \\\hline
\end{tabular}
\end{table*}

The predictability period mainly depends on the close approaches with the Jupiter. As we can see in table 2 the most predictable orbits have objects which avoid close encounters with this planet, i.e. \cu, \gh, \qq\ and \aee. The three identified horseshoe librators \ql, \sa\ and \Larsen\ have short predictability period mostly due to the specific dynamic of horseshoe orbit. When an object is moving on a HS orbit, it has large amplitude of libration and experiences close multiple encounters with the planet (even if its eccentricity is small). This generates instability of the object's orbit and one can expect that such an asteroid is expelled from the co-orbital region after a few HS libration periods (see also discussion in Stacey and Connors (2008), section 4.1 therein). Indeed, according to our analysis these three objects execute a few HS loops (in the case of \ql\ and \Larsen\ the HS state is interrupted by QS behavior) before leaving this state, as well as the planet's co-orbital region. The comets \ar\ and \wc\ do not experiences a close encounters with Jupiter. On the other hand, these objects can close approach to Mars. Additionally, these comets have very short observational intervals. For these reasons the predictability period of these object is short. Asteroids \hr\ and \ug\ experience close encounters with Jupiter and they have relatively short predictability time interval. These objects also remain in the co-orbital region at most a few hundred of years. We conclude that the objects that experience close encounters with the Jupiter may be delivered into the co-orbital region from different parts of the Solar System, for example from either the Trojan area of external part of the System and they leave the co-orbital region relatively quickly (within a few hundred years). Kortenkamp and Joseph (2011) found that the Jupiter Trojans were able to migrate to QS orbits during the late stages of migration of the giant planets, however, it seems that current QS of this planet entered into this state later. The main reason is that the mechanism of transforming TP orbit into QS orbit was not efficient (about 0.2\% of the escaping Jupiter Trojan particles entered in QS after leaving the Trojan region within a few milion of years) and the QS of Jupiter can persist in this state at most $10^7$ yrs (Wiegert et al.  2000). Thus, the present QS entered into this state likely a few millions year ago or less.

We have shown that \gh\ is currently a QS of Jupiter for about 200 years and it leaves this regime about 2030 transiting into TP orbit. In general, this object stays within the co-orbital region (all VOs remain in this area) and it jumps between QS and TP types of orbits. Similar dynamics demonstrates \gh, however, according to our calculations after 3900 it will be moving in TP-QS-TP orbit with large amplitude of the guiding center. This object has been observed for a short time interval of about 1.33 yr but it exhibits unexpectedly very long predictability period. The orbit of \gh\ is predictable to $\sim$4500. After that time the object is still locked in a 1:1 mean motion resonance with Jupiter to the end of integration.

In this paper the detailed analysis of the four QS objects, i.e. two asteroids (\qq\ and \aee) and two comets (\ar\ and \wc) found by Kinoshita and Nakai (2007) have been presented. For both asteroids our results fully agree with the conclusions
drawn by Kinoshita and Nakai (2007). This is especially worth emphazing
because asteroid 2001 QQ199 was observed after the year 2007, thus for this object we used significantly richer observational material than Kinoshita and Nakai (2007). Since the NG effects may affect the dynamical behavior of comets \ars\ and \wcs\, we decided  to integrate their orbits only 1000 years backward and forward. Within this period of integration all VOs of the first comet does not diverge significantly and all VOs shows QS behavior. In the case of \wcs\ we have shown that this object is a QS of Jupiter to the year 2500 at least. After that time the orbital elements of this comet starts to diverge and according to presented calculations the transitions between QS and TP orbit can take place. The main reasons why the predictability period of this object is very short seems to be that the comet has only about 0.5 yr interval of observations and the large eccentricity. Although \wcs\ does not pass close to Jupiter (all encounters are over 2.0 a.u.), it is able to experience very close encounters with Mars and the Earth due to the very eccentric orbit (e=0.68).

We have confirmed the results of Karlsson (2004) that \hr\ will be temporarily captured by Jupiter on a QS orbit near 2050 and that the object \ug\ will be captured into co-orbital region of Jupiter around 2045, however, in contrary to his predictions, this asteroid is not co-orbital of Jupiter, presently. Its guiding center encloses the origin but this object will not a QS of Jupiter. Currently, these objects do not librate in a 1:1 mean motion resonance with Jupiter. This behavior causes that \hr\ and \ug\ repeatedly are coming close to Jupiter and therefore their predictability period is of order of a few hundred years.

\newpage

\Acknow{Part of numerical calculations was performed using the numerical
orbital package developed by Professor Grzegorz Sitarski and the
Solar System Dynamics \& Planetology Group at SRC PAS.}

\end{document}